\providecommand{\exlude}[1]{}

\documentclass[10pt, twocolumn, showpacs, aps, pra, superscriptaddress,
               preprintnumbers, longbibliography, nofootinbib]{revtex4-1}

\usepackage{arXiv}   % Hacks to get arXiv working with their decade old tex distro.

\usepackage[utf8]{inputenc}     % Allow unicode characters in document like ł
\usepackage{graphicx}
\usepackage{my_acronyms}
\usepackage{hyperref}
\usepackage[version=4]{mhchem}
\usepackage{siunitx}
\usepackage{calc}

\graphicspath{{./}{figures/}}       % Allow figures to be kept separately

\let\Re\relax
\DeclareMathOperator{\Re}{Re}
\newcommand{\abs}[1]{\lvert#1\rvert}

\begin{document}
\newcommand{\todo}[1]{}

\preprint{NT@UW-16-10}
\preprint{INT-PUB-16-026}

\title{Towards Quantum Turbulence in Cold Atomic Fermionic Superfluids}

\author{Aurel Bulgac}
\email{bulgac@uw.edu}
\affiliation{Department of Physics, University of Washington, %
  Seattle,  WA 98105--1560, USA}

\author{Michael McNeil Forbes}
\email{michael.forbes@wsu.edu}
\affiliation{Department of Physics and Astronomy, %
  Washington State University, %
  Pullman,  WA 99164--2814, USA}
\affiliation{Department of Physics, University of Washington, %
  Seattle,  WA 98105--1560, USA}

\author{Gabriel Wlaz\l{}owski}
\email{gabrielw@if.pw.edu.pl}
\affiliation{Faculty of Physics, %
  Warsaw University of Technology, %
  Ulica Koszykowa 75, 00--662 Warsaw, POLAND}
\affiliation{Department of Physics, University of Washington, %
  Seattle,  WA 98105--1560, USA}

\begin{abstract}
  \noindent
  Fermionic superfluids provide a new realization of quantum turbulence,
  accessible to both experiment and theory, yet relevant to phenomena from both
  cold atoms to nuclear astrophysics.  In particular, the strongly interacting
  Fermi gas realized in cold-atom experiments is closely related to dilute
  neutron matter in neutron star crusts.  Unlike the liquid superfluids
  \ce{^4He} (bosons) and \ce{^3He} (fermions), where quantum turbulence has
  been studied in laboratory for decades, superfluid Fermi gases stand apart
  for a number of reasons.  They admit a rather reliable theoretical
  description based on \gls{DFT} called the \gls{TDSLDA} that describes both
  static and dynamic phenomena.
  %
  % This is a natural extension of the Kohn and Sham implementation of \gls{DFT}
  % to time-dependent superfluid systems. 
  %
  Cold atom experiments demonstrate exquisite control over particle number,
  spin polarization, density, temperature, and interaction
  strength. Topological defects such as domain walls and quantized vortices,
  which lie at the heart of quantum turbulence, can be created and manipulated
  with time-dependent external potentials, and agree with the time-dependent
  theoretical techniques.  While similar experimental and theoretical control
  exists for weakly interacting Bose gases, the unitary Fermi gas is strongly
  interacting. The resulting vortex line density is extremely high, and quantum
  turbulence may thus be realized in small systems where classical turbulence
  is suppressed.
  Fermi gases also permit the study of exotic superfluid phenomena such as the
  \gls{LOFF} pairing mechanism for polarized superfluids which may give rise to
  3D supersolids, and a pseudo-gap at finite temperatures that might affect the
  regime of classical turbulence.  The dynamics associated with these phenomena
  has only started to be explored. Finally, superfluid mixtures have recently
  been realized, providing experimental access to phenomena like
  Andreev-Bashkin entrainment predicted decades ago.  Superfluid Fermi gases
  thus provide a rich forum for addressing phenomena related to quantum
  turbulence with applications ranging from terrestrial superfluidity to
  astrophysical dynamics in neutron stars.
\end{abstract}

\glsreset{LOFF}

%\keywords{}

%\submitto{\jpb}

\date{\today}

\maketitle

\begin{figure*}[t]
  \includegraphics[width=0.4\textwidth]{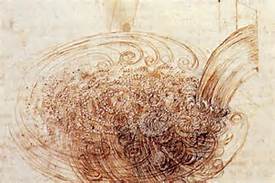}%
  \hspace{2em}%
  \includegraphics[width=0.4\textwidth]{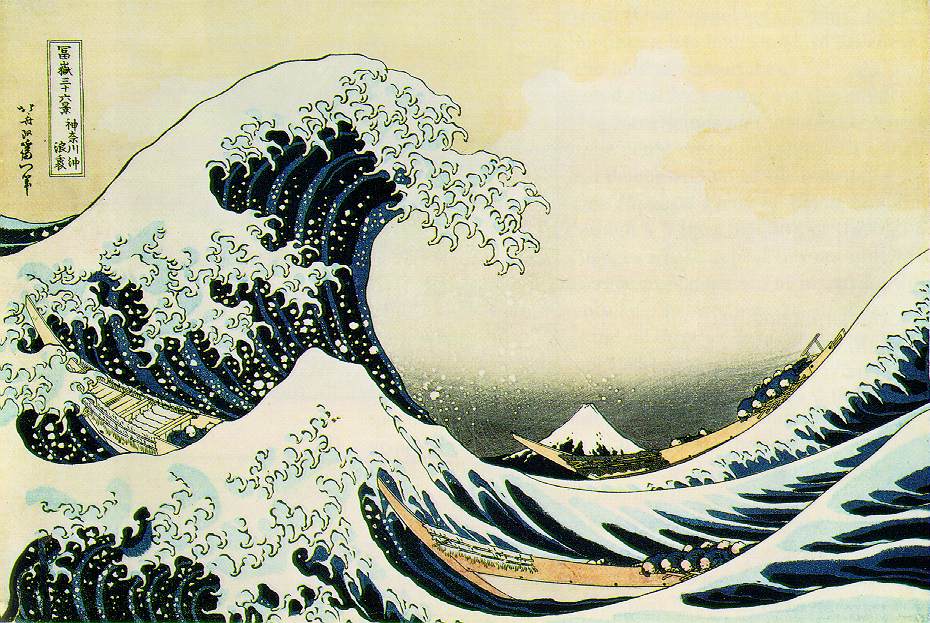}%
  \caption{Works by
    Leonardo da Vinci (left) and Katsushika Hokusai (right) illustrating 
    the complexity and the captivating beauty of the turbulent flow.}%\vspace{-4mm}
  \label{FIG:paintings}
\end{figure*}

\noindent
Hydrodynamic turbulence has fascinated thinkers and artists since medieval
times, and many authors use in their presentations the works of Leonardo da
Vinci and Katsushika Hokusai to illustrate the complexity and beauty of
turbulent flow (see Fig.~\ref{FIG:paintings}).  Classical flow is characterized
by its Reynolds number $\Re = \rho vL/\eta$,
where $\rho$
is the fluid mass density, $L$
the length of the system, $v$
the flow velocity, and $\eta$
the shear viscosity. In three-dimensional fluids, turbulent motion is
observed only if the Reynolds number is larger than
$\Re \gtrsim 10^4$~\cite{Lamb:1945,LL6:1966}.
Since superfluids have no shear viscosity and behave as a perfect classical fluid
at zero temperature, turbulence was naively not expected to emerge at
subcritical velocities.  However, in 1955 Feynman conjectured that the
quantized vortices~\cite{Feynman:1955} predicted by him and
Onsager~\cite{Onsager:1949} could lead to a new type of turbulence: quantum
turbulence.  A tangle of vortex lines, which visually appears as a bunch of
tangled spaghetti, will evolve in time in a rather chaotic fashion, cross and
recombine in similar fashion to strands of DNA and lead to quantum
turbulence. At finite temperatures, but below the critical temperature, a
normal component is also present and the normal and superfluid components 
interact.  
In the literature, the term quantum turbulence
also refers to the dynamics of finite-temperature superfluids where the vortices
in the superfluid component interacts with a classically turbulent normal component.

%&&&&&&&&&&&&&&&&&&&&&&&&&&&&&&&&&&&&&&&&&&&&&&&&&&
%&&&&&&&&&&&&&&&&&&&&&&&&&&&&&&&&&&&&&&&&&&&&&&&&&& 

Until recently studies of quantum turbulence have been performed mostly in
liquid \ce{^4He} and \ce{^3He}~\cite{Barenghi:2001, Vinen:2006,
  Vinen:2007, Vinen:2010, Paoletti:2011, Barenghi:2014}, and lately also in
\glspl{BEC} of cold atoms~\cite{Bagnato:2016}. Both liquid \ce{^4He} and
\ce{^3He} are strongly interacting systems of bosons and fermions respectively,
but unfortunately a microscopic theory describing their dynamics does not
exists yet. On the other hand the \gls{BEC} of cold atoms can be described
quite accurately either by the \gls{GPE}~\cite{Gross:1961, Pitaevskii:1961} at
very low temperatures, when the fraction of the normal component is negligible,
or by the theory of \gls{ZNG}~\cite{Zaremba:1999}, in which the dynamics and
the coupling between the normal and superfluid components are accounted for.
There exists also stochastic extensions of the \gls{GPE}~\cite{Gardiner:2002,
  Gardiner:2003}, in which two types of modes are introduced, slow quantum
modes describes by the GP equation, and the fast modes modeled stochastically,
separated by a cutoff energy. Unfortunately, these stochastic extensions appear
to generate results which are dependent on the choice of the cutoff
energy~\cite{Davis:pc}. The \gls{BEC} of cold gases are systems of weakly
interacting atoms, which is the main reason why the derivation of either the
\gls{GPE} or the \gls{ZNG} framework were possible.

There exist a number of alternative phenomenological approaches to study
superfluid dynamics.  As noted by Sato and Packard~\cite{Sato:2012x}, in the
case of Bose superfluids ``two phenomenological theories explain almost all
experiments'': the two-fluid hydrodynamics due to Tisza and
Landau~\cite{Tisza:1938, Tisza:1940, Tisza:1940a, Landau:1941, Landau:1941a,
  Khalatnikov:2000}, and the ``complementary view provided by Fritz London,
Lars Onsager, and Richard Feynman, [that] treats the superfluid as a
macroscopic quantum state.''  A shortcoming of the two-fluid hydrodynamics,
which is ``essentially thermodynamics''~\cite{Sato:2012x}, is the absence of
the Planck's constant and the corresponding lack of quantized vortices. Vortex
quantization was later implemented ``by hand'' when
needed~\cite{Khalatnikov:2000}, or by using the \gls{GPE}.  Another remedy is
the widely-used filament model~\cite{Schwarz:1988, Tsubota:2013, Hanninen:2014}
in which quantized vortices and their crossing and reconnections are introduced
\textit{ad hoc} using the Biot-Savart law as a recipe to construct the
resulting velocity from the vortex lines.

Unlike \glspl{BEC}, the \gls{UFG} or a Fermi gas near unitarity -- a gas of
cold fermionic atoms with interactions tuned at or near a Feshbach resonance --
is strongly interacting, and exhibits a large pairing gap of the order of the
Fermi energy, a short coherence length of the order of the average
interparticle separation, and a high critical velocity of the order of the
Fermi velocity~\cite{CCPS:2003, CPCS:2004, giorgini-2007, Zwerger:2011,
  Sensarma:2006, Combescot:2006}.  The \gls{UFG} shares many similarities with
the dilute neutron matter, which is present in the skin of neutron-rich nuclei,
and in the crust of neutron stars where neutron-rich nuclei form a Coulomb
lattice immersed in a neutron superfluid. This motivated the initial
independent microscopic studies of the \gls{UFG} by nuclear
theorists~\cite{baker00:_mbx_chall_compet, Baker:1999, CCPS:2003}.\footnote{The
  The Many-Body Challenge Problem (MBX) formulated by G.~F.~Bertsch in 1999.
  See also~\cite{baker00:_mbx_chall_compet, Baker:1999}.} A system of fermions
with zero-range interaction and infinite scattering length should have a ground
state energy and properties determined by their only existing dimensional
scale, namely their density, similarly to a free Fermi gas. Theoretical
\gls{QMC} studies of the Fermi gas near unitarity have achieved percent-level
accuracy~\cite{Astrakharchik:2004, Carlson:2011, FGG:2010, Forbes:2012} and
agree with experiment~\cite{Ku:2011} at an accuracy better than 1\% for the
energy per particle, and a few percent for the magnitude of the pairing
gap~\cite{Carlson;Reddy:2008-04, Schirotzek;Shin;Schunck;Ketterle:2008-08}.

A microscopic theoretical framework capable of describing quantum turbulence in
fermionic superfluids and implementable in realistic calculations has became
possible only recently.  Two factors played an crucial role: i)~the development
and validation against experiment of an appropriate microscopic framework for
the structure and dynamics of fermionic superfluids, and ii)~the implementation
of this framework using sophisticated numerical algorithms that fully utilize
the advanced capabilities of modern leadership class computers, such as
Titan.\footnote{\url{https://olcf.ornl.gov/computing-resources/titan-cray-xk7/}. The
  ensemble of all \num{18688} \glspl{GPU} on Titan delivers about 90\% of the
  computing power, while the rest of \num{299008} \glspl{CPU} provide the rest
  of about 10\%. A single \gls{GPU} provides the same number of floating point
  operations as approximately 134 \glspl{CPU}.\label{fn:Titan}}

Quantum Monte Carlo algorithms cannot be used to describe turbulent dynamics
and the only available theoretical candidate with a microscopic underpinning is
\gls{TDDFT}~\cite{HK:1964,Kohn:1965fk,Dreizler:1990lr,Gross:2006} extended to
superfluid systems. Such an extension of \gls{TDDFT} performed in the spirit of
the Kohn and Sham~\cite{Kohn:1965fk} is known as the
\gls{TDSLDA}~\cite{Bulgac:2011, Bulgac:2013b}.  This is in principle an exact
approach~\cite{Drut:2010kx}, in the same sense as the initial formulation of
\gls{DFT} by Hohenberg and Kohn~\cite{HK:1964}.  The \gls{SLDA}
and the Schr\"odinger equation description of a given system should be
indistinguishable at the one-body level. One can prove mathematically that an
\gls{EDF} for a given system exists~\cite{HK:1964}, but unfortunately no
mathematical recipe has been ever devised for constructing this \gls{EDF}. In
the case of the \gls{UFG}, however, the structure of the functional can be
fixed with high accuracy by rather general requirements: dimensional arguments
require that it can depend only on the fermion mass $m$,
Planck's constant $\hbar$,
the number density $n(\vec{r})$,
the kinetic energy density $\tau(\vec{r})$,
the anomalous density $\nu(\vec{r})$,
and the current density $\vec{j}(\vec{r})$.
The inclusion of the anomalous density is required in order to be able to
disentangle the normal from the superfluid phases, and the presence of currents
is necessary to put in evidence matter flow.  These must be combined in an
\gls{EDF} that preserves Galilean covariance and appropriate symmetries
(parity, translation, rotation, gauge, etc.).  As in the implementation by Kohn
and Sham~\cite{Kohn:1965fk} of the underlying \gls{DFT}~\cite{Kohn:1965fk}, the
kinetic energy density largely accounts for gradient effects, and additional
corrections appear to be rather small~\cite{Bulgac:2007a, Forbes:2012,
  Forbes:2012a} in case of the \gls{UFG}.  The resulting (unregulated and
unrenormalized) \gls{EDF} for an unpolarized \gls{UFG} reads:
\begin{subequations}
  \begin{multline}
    \varepsilon(\vec{r}) 
    = 
    \frac{\hbar^2}{m}
    \Biggl[\alpha  \frac{\tau(\vec{r})}{2} 
    + \gamma \frac{\abs{\nu(\vec{r})}^2}{n^{1/3}(\vec{r})} 
    +\beta \frac{3(3\pi^2)^{2/3}n^{5/3}(\vec{r})}{5}\\
    - (\alpha-1)\frac{\vec{j}^2(\vec{r})}{2n(\vec{r})}\Biggr], \label{eq:edf}
  \end{multline}
  where
  \begin{align}
    n(\vec{r}) &= 2\sum_{0<E_k<E_c}|v_k(\vec{r})|^2,\\
    \tau(\vec{r}) &= 2\sum_{0<E_k<E_c}|v_k(\vec{r})|^2,\\
    \nu(\vec{r}) &= \sum_{0<E_k<E_c}u_k(\vec{r})v_k^*(\vec{r}),\\
    \vec{j}(\vec{r}) &= 2\,\mathrm{Im}\sum_{0<E_k<E_c}v_k^*(\vec{r})\vec{\nabla}v_k(\vec{r}),
  \end{align}
\end{subequations}
and where the sums are taken up to a specified energy cutoff $E_c$
in the quasiparticle energy $E_k$, see below.
This \gls{EDF} is written for simplicity in its unregulated form, and the full
theory requires regularization to properly deal with the kinetic and anomalous
densities which diverge as functions of $E_c$.
The regularization and renormalization process is well established and described
in~\cite{BY:2002fk,Bulgac:2011}. Terms in this functional have very simple physical
meaning. The first represents kinetic energy contribution, while the second is due to the
pairing interaction -- these two terms must enter in a specific combination
in order to assure the theory can be regularized.  The third term represents
the normal interaction energy, and the last term restores Galilean covariance of the
functional.  The dimensionless constants $\alpha$, $\beta$, and $\gamma$
are determined from independent \gls{QMC} calculations of the uniform
\gls{UFG}~\cite{CCPS:2003, Carlson:2005kg, Carlson;Reddy:2008-04,
  Carlson:2011}.  The functional has also been extended to spin-polarized
case. (For an extensive review, see~\cite{Bulgac:2011}.)  In the stationary
case, the quasi-particle wave-functions satisfy the Bogoliubov-de Gennes like
equations
\small
\begin{equation}
  \begin{pmatrix}
    (h(\vec{r})-\mu) & \Delta(\vec{r}) \\ 
    \Delta^*(\vec{r}) &  -(h^*(\vec{r})-\mu)
  \end{pmatrix}
  \begin{pmatrix}
    u_k(\vec{r})\\ 
    v_k(\vec{r})
  \end{pmatrix}
  = E_k  \begin{pmatrix} 
    u_k(\vec{r})\\
    v_k(\vec{r})
  \end{pmatrix},
  \label{eq:slda}
\end{equation}
\normalsize
where
\begin{align}
  h(\vec{r}) 
  &= \frac{\delta \varepsilon(\vec{r})}{\delta n(\vec{r})} +
    V_{\text{ext}}(\vec{r}), 
  &
    \Delta(\vec{r}) 
  &= \frac{\delta \varepsilon(\vec{r})}{\delta \nu^*(\vec{r})},
\end{align}
and where $V_{\text{ext}}(\vec{r})$ is an arbitrary external potential and
$\mu$ is the chemical potential.

The parameters of the \gls{SLDA} fixed by homogeneous systems have been further
validated against solutions of the Schr\"odinger equation for inhomogeneous
systems, and found to agree on the few percent level or better with \gls{QMC}
results for almost 100 systems~\cite{Blume;Stecher;Greene:2007-12,
  D.Blume:2008-05, Stecher;Greene;Blume:2008-12} including both polarized and
unpolarized systems, as well as superfluid and normal systems~\cite{Bulgac:2011}.
This step is necessary to demonstrate that the Schr\"odinger equation and \gls{SLDA}
for the \gls{UFG} deliver identical descriptions of various quantum systems as required by
the general \gls{DFT} theorem~\cite{HK:1964,Kohn:1965fk}.

In the case of time-dependent problems, the external potential could become
time-dependent (an external stirrer) and the equations for the quasi-particle
wave-functions for a spin unpolarized system become
\small
\begin{equation*}
  i\hbar\begin{pmatrix}
    \dot{u}_k(\vec{r},t)\\
    \dot{v}_k(\vec{r},t)
  \end{pmatrix}
  = 
  \begin{pmatrix}
    (h(\vec{r},t)-\mu) & \Delta(\vec{r},t) \\ 
    \Delta^*(\vec{r},t) &  -(h^*(\vec{r},t)-\mu)
  \end{pmatrix}
  \begin{pmatrix}
    u_k(\vec{r},t)\\
    v_k(\vec{r},t)
  \end{pmatrix}.
\end{equation*}
\normalsize
The structure of these \gls{TDSLDA} equations and the static \gls{SLDA}
equations (\ref{eq:slda}) illustrates the numerical complexity of the problem:
one must solve them for all quasi-particle states up to cutoff energy $E_c$.
The number of such equations is in principle infinite and they must be
accurately discretized, but even after discretization one ends up with the
order of $N_xN_yN_z$
quasi-particle wave-functions depending on $N_xN_yN_z$
spatial lattice points, where $N_x$,
$N_y$,
and $N_z$
are the number of spatial lattice points in each spatial direction.  For
typical problems studied so far, the number of quasi-particle wave-functions
ranges from several tens of thousands to a fraction of a million, leading
correspondingly to twice the number of nonlinear coupled complex time-dependent
three-dimensional partial differential equations. The solution of these coupled
equations requires leadership class computers with hardware accelerators such
as \gls{GPU}.  In practical applications (due to the numerical complexity and
memory requirements) thousands of \glspl{GPU} are required, and presently
the supercomputer Titan at Oak Ridge National Laboratory (the largest supercomputer
currently in US) is one of the few than can handle this type of
calculations (see footnote~\ref{fn:Titan}).

The dynamics of quantized vortices, their crossing, and recombination, requires
a full solution in 3D of the \gls{TDSLDA} equations for a large number of
time-steps (hundreds of thousands to millions).  This makes the use of
leadership class computers an inextricable part of the solution of these kind
of problems.  Both the development of a validated theoretical microscopic
framework and the availability of supercomputers were crucial developments
in making the study of fermionic superfluid possible.

One of the first illustrations of the power of the \gls{TDSLDA} was to show for
the first time in literature how various type of quantized vortices can be
generated dynamically by stirring a cloud, and how these vortices propagate,
cross, recombine, and eventually disappear from the system by interacting with
the boundaries~\cite{Bulgac:2011b}. It was also demonstrated that the
superfluid system can become normal at relatively large stirring velocities,
but also, somewhat unexpectedly, that a system stirred at a velocity nominally
larger than the speed of sound can remain superfluid.  This behavior is
different than in superfluid \ce{^4He} or
\ce{^3He}, and results from the high compressibility
of the \gls{UFG}, where the speed of sound can locally increase and stabilize the
superfluid phase.

Solutions of the time-dependent Schr\"odinger equation for many-fermion systems
are hard to come by.  There are a few cases where analytical solutions can be
obtained~\cite{Yuzbashyan:2006a, Yuzbashyan:2006, Dzero:2015} and some have
been compared to the corresponding solutions of the similar \gls{TDSLDA}
systems~\cite{Bulgac:2009}, corresponding to excitations of the Anderson-Higgs
modes and related excitations that await experimental confirmation. The
Anderson-Higgs mode is rather spectacular and has some similarities with a
plasmon mode in the sense that it has a finite excitation energy for zero
momentum.  The Anderson-Higgs mode is however not a number density oscillation,
the number density remains constant in a homogeneous system and only the amplitude of the pairing gap
undergoes large slow anharmonic oscillations in time.

There are a number of non-trivial non-equilibrium phenomena studied
experimentally and these results can be confronted with the \gls{TDSLDA}
predictions. One such experiment addresses the excitation of so called quantum
shock-waves~\cite{Joseph:2011}.  In a shock-wave, properties of the system vary
drastically across the shock-wave front, and have distinct and almost
time-independent values on either side of the shock-wave front. Classically
shock-waves owe their existence to dissipative effects
and their interplay with dispersive effects~\cite{Whitham:1974}. Dissipation,
however, is strongly inhibited in quantum systems at low temperatures and the
nature of quantum shock-waves is not obvious~\cite{El:2016}.  The \gls{TDSLDA}
is able to describe rather accurately the existence and
the shape of quantum shock-waves observed in experiments, without the need to
introduce any dissipative effects~\cite{Bulgac:2011c}.  The shock-wave front
propagates at supersonic speeds, as expected.  Moreover, the \gls{TDSLDA} also
predicts the excitation of domain walls in the wake of the shock-wave front, with
speeds lower than the speed of sound.  These domain walls are topological
excitations of the pairing field, which has a negative but almost constant value
on one side of the domain wall and positive value on the other (defined up to an 
obvious gauge transformation). The accuracy of
the experiments on quantum shock-waves performed so far is insufficient to put
in evidence these domain walls, but at a hydrodynamic level, these
experiments are well described by a \gls{GPE}-like theory tuned to the
\gls{UFG} equation of state~\cite{Ancilotto:2012}.  
In such a \gls{GPE}-like description one cannot distinguish 
between density and pairing order parameters and such an approach
will also fail to describe the pair-breaking excitations and, moreover,  the
Anderson-Higgs mode as well, when the density is constant, but the pairing
field oscillates. In a \gls{GPE} approach, the number density cannot be disentangled
from the superfluid order parameter.

\begin{figure*}[tb]
  \includegraphics[angle=-90,origin=br,width=0.25\textwidth]{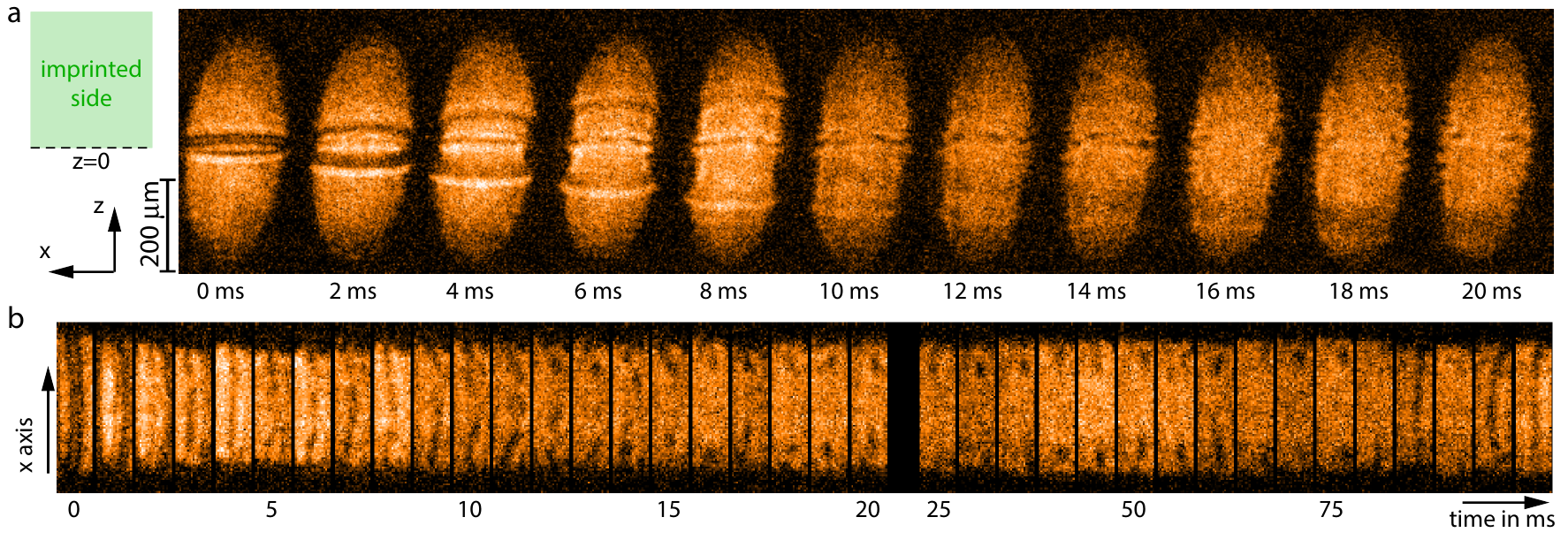}%
  \includegraphics[width=0.4\textwidth]{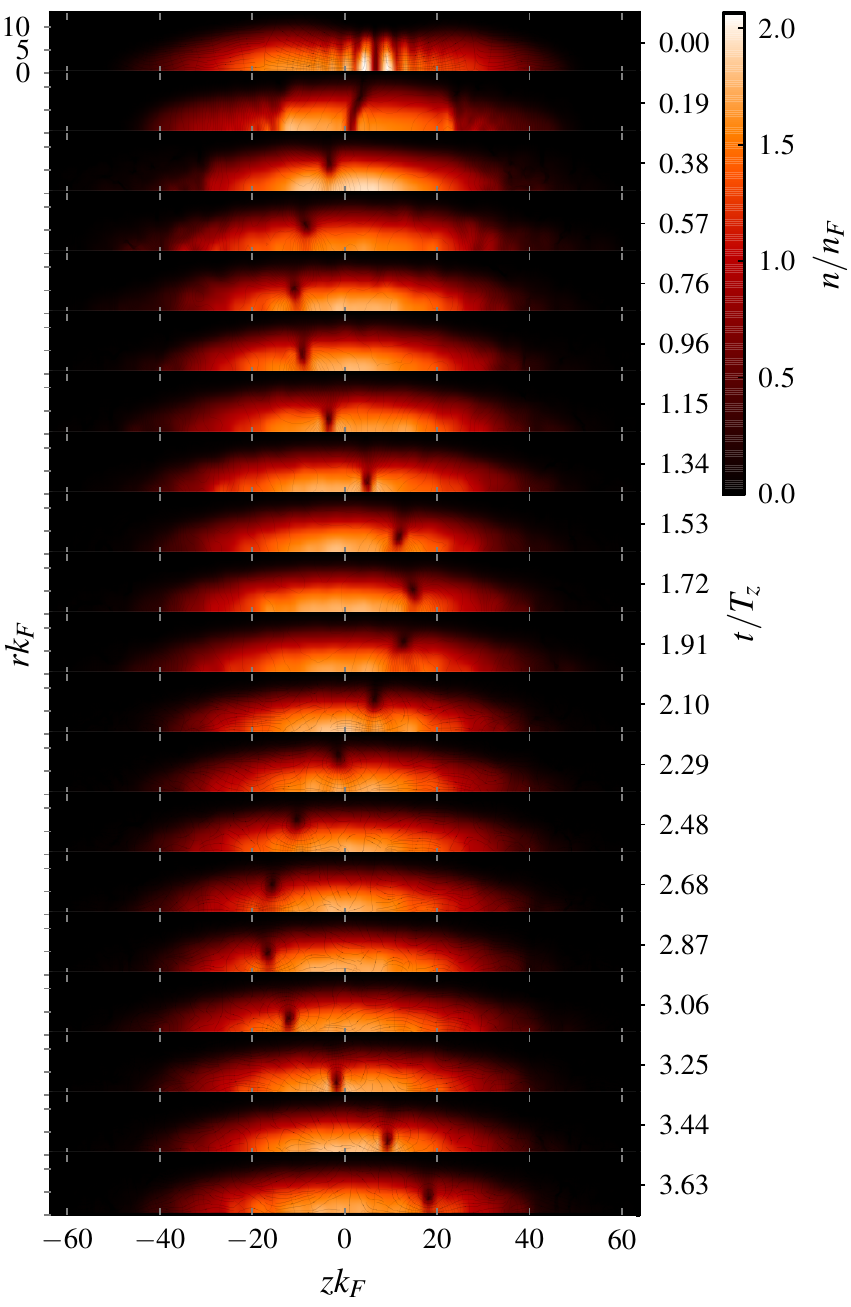}
  \caption{\label{fig:densities} Evolution of a phase imprinted
    domain wall in a harmonically trapped unitary Fermi gas.  The images show
    the density along a central slice as a function of time from top to bottom.
    The left panel shows the evolution of the experiment~\cite{Ku:2015}
    demonstrating the evolution of the initial domain wall into vortex rings
    and finally solitonic vortex segments.  The right images show the numerical
    \gls{TDSLDA} evolution of a similar (but smaller) system in an axially
    symmetric geometry from~\cite{Bulgac:2013d} demonstrating the evolution of
    the initial domain wall into a vortex ring which oscillates. The
    \gls{TDSLDA} results appear to extrapolate correctly to larger
    systems. (Images reproduced with permission.)  }
\end{figure*}

\begin{figure*}[htb]
  \def\tmp#1{\includegraphics[width=0.12\textwidth]{#1}}%
  \tmp{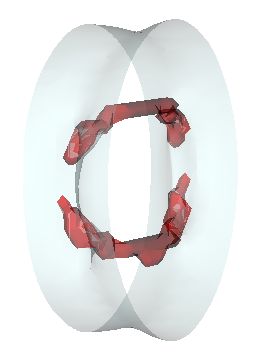}%
  \tmp{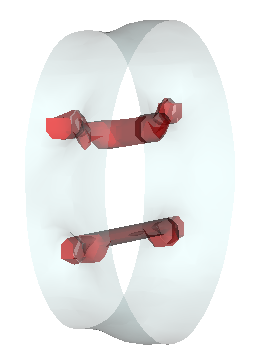}%
  \tmp{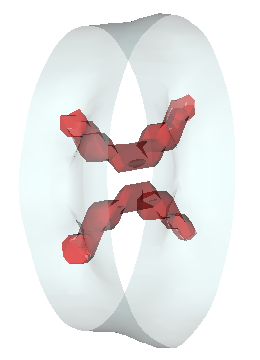}%
  \tmp{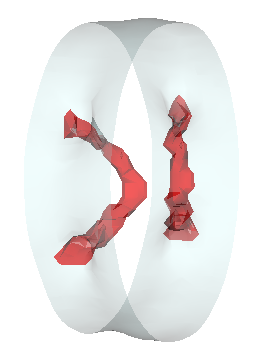}%
  \tmp{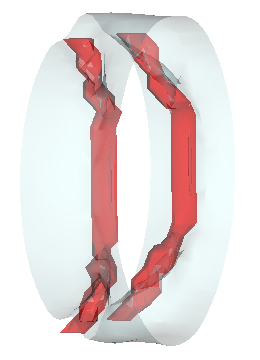}\\
  \hspace{10mm}%
  \tmp{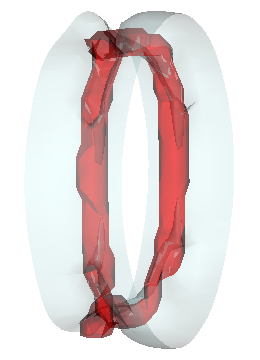}%
  \tmp{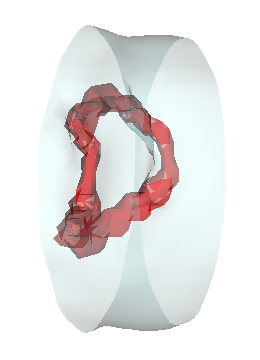}%
  \tmp{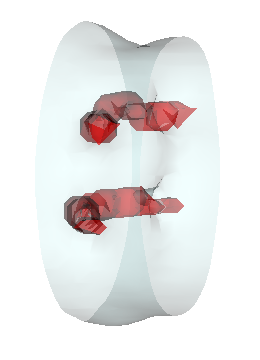}%
  \tmp{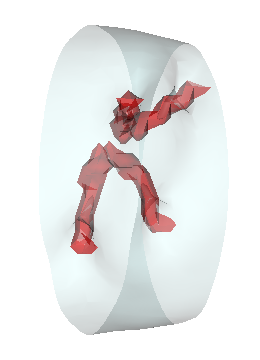}%
  \tmp{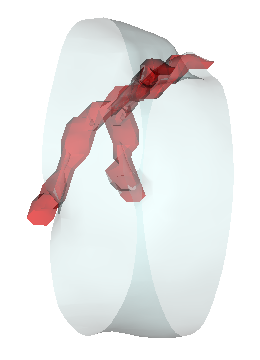}%
  \tmp{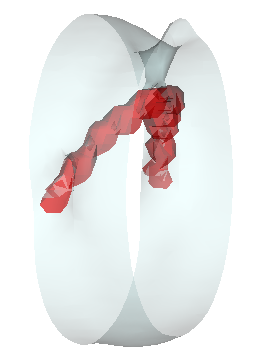}
  \let\tmp\relax
  \caption{\label{fig:vortexcross} Vortex reconnections
    transforming a vortex ring into a vortex line.  An initial vortex ring, 
    touches the boundaries and coverts into two vortex lines. The first reconnection
    occurs closer to one edge of the cloud and ends with formation of a new
    vortex ring (top row).  The new vortex ring moves to other side of the cloud. It is deformed and undergoes
    asymmetric deformation into two vortex lines, one that is absorbed by the boundary and one
    that survives (bottom line).}
\end{figure*}

Another strong validation of the correctness of the \gls{TDSLDA} was the
unambiguous and correct identification of the nature of the excitation observed
by the MIT group, identified by the authors as a ``heavy
soliton''~\cite{Yefsah:2013}. Using the \gls{TDSLDA} framework (right side of Fig~\ref{fig:densities}), we demonstrated
how an initially formed domain wall would evolve rapidly into a vortex
ring~\cite{Bulgac:2013d} in the elongated axially symmetric traps described in
the initial paper~\cite{Yefsah:2013}.  Vortices in the \gls{UFG} have a small
core and are difficult to image in experiment.  For this reason the MIT group
devised an ingenious imaging procedure to visualization the excitations.  This
procedure significantly altered the size of the excitations, originally leading
to the incorrect identification of these objects as thick heavy solitons,
instead of the thin vortices and vortex rings that were actually produced.
Using \gls{TDDFT} techniques we theoretically simulated this experimental
visualization protocol, demonstrating that vortex rings expand to a size
consistent with the thick objects observed in the experiment.  The MIT group
subsequently refined their imaging procedure, developing a new tomographic
imaging technique capable of isolating portions of the excitations, thereby
confirming that the imprinted domain wall indeed evolves into a vortex ring,
and ultimately a vortex segment when axial symmetry is broken~\cite{Ku:2014}.
In this subsequent paper it
was revealed that the axial symmetry of the trap was spoiled by gravity.
\gls{TDSLDA} simulation of these asymmetric traps confirmed the evolution of
the vortex ring into a vortex segment as the ring collides with the trap
boundaries~\cite{Wlazlowski:2015}. Similar processes were first demonstrated in
dynamical simulations of the \gls{UFG}~\cite{Bulgac:2011b}.  Thus, all of the
qualitative aspects of these dynamic experiments have been reproduced by the
\gls{TDSLDA} framework.

Interestingly, the \gls{TDSLDA} predicts the crossing and reconnection of
vortex lines in the intermediate stage between the vortex ring and the vortex
line as demonstrated in Fig.~\ref{fig:vortexcross}.  This prediction still
awaits experimental verification.  The latest MIT paper~\cite{Ku:2015} (see
Fig.~\ref{fig:densities}) suggests the existence of exotic structures in the
intermediate stage (between domain wall and vortex line state), a combination
of a vortex ring and one or more vortex lines, called a
$\Phi$-soliton~\cite{Munoz-Mateo:2014}.
The existence of a $\Phi$-soliton
needs further experimental investigation as no dynamical mechanism for their
generation, from a domain wall to a $\Phi$-soliton
state, has yet been demonstrated in any simulation.  Quantitative validation of
the \gls{TDSLDA} is still required, and will need accurate experimental
measurements with improved imaging capable of resolving such states in systems
small enough to accurately simulate ($\sim 1000$
fermions on present computers).

\begin{figure*}[htb]
  \def\tmp#1#2{\includegraphics[width=0.5\textwidth]{#1}%
               \includegraphics[width=0.3\textwidth, trim=150 20 0 0, clip]{#2}}
  \tmp{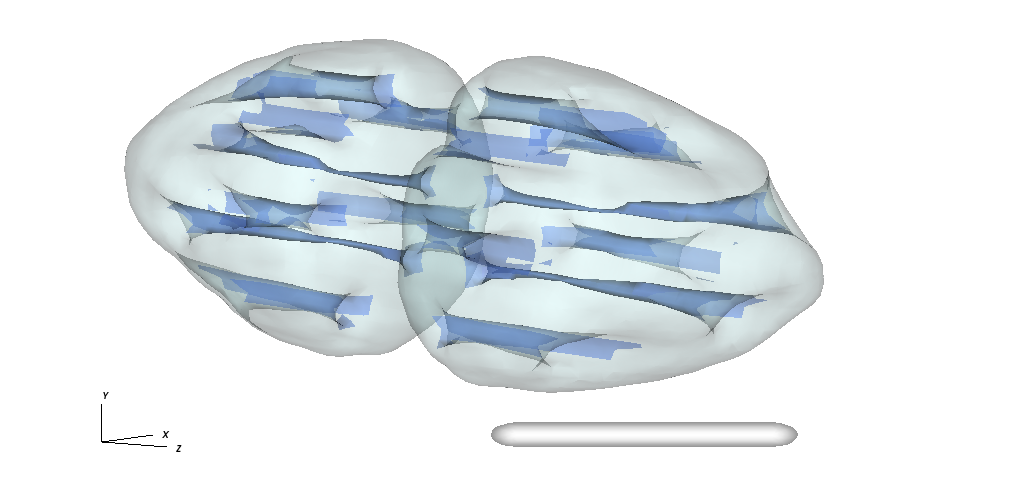}{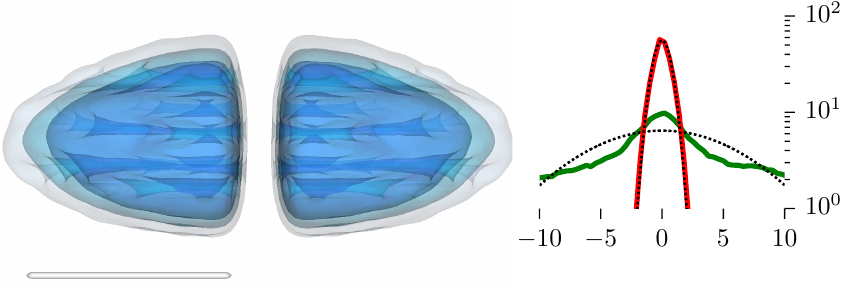}\\
  \tmp{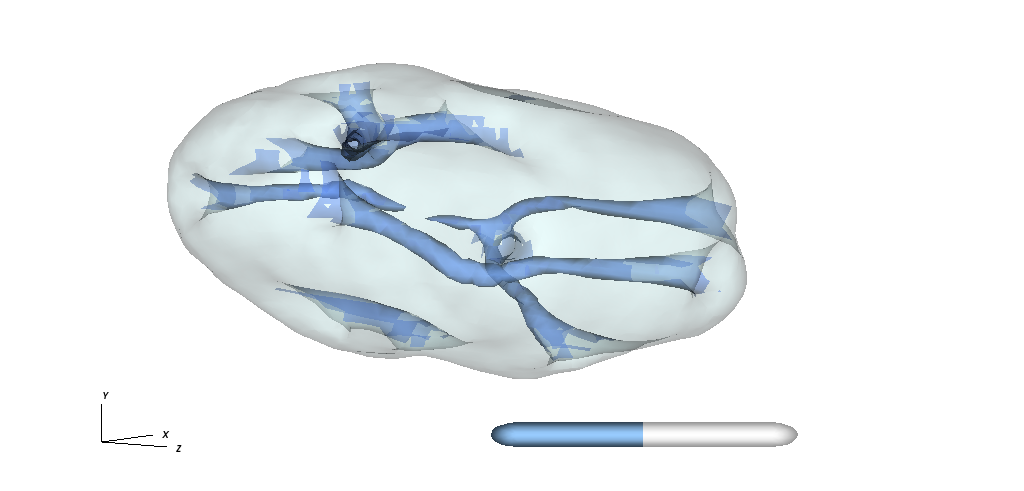}{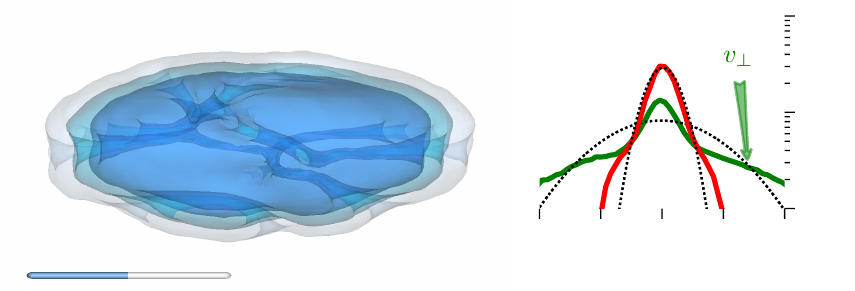}\\
  \tmp{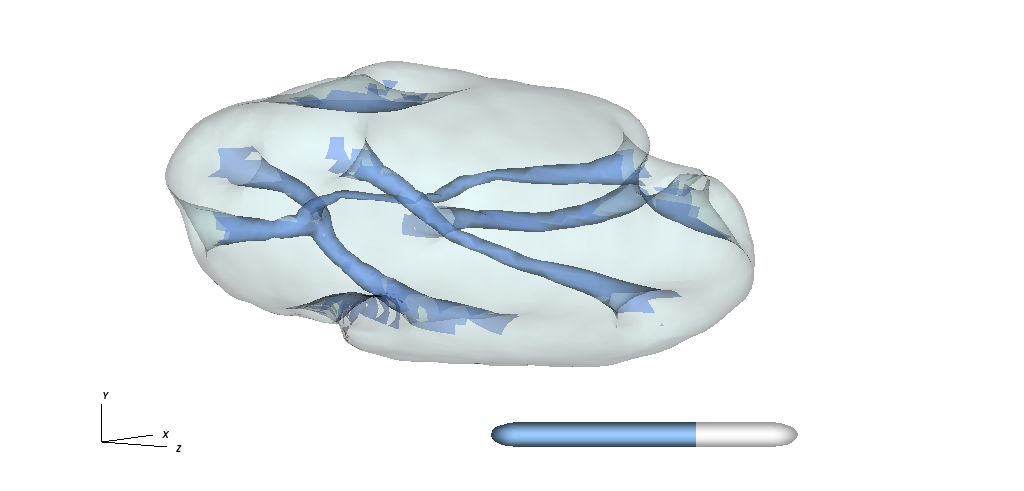}{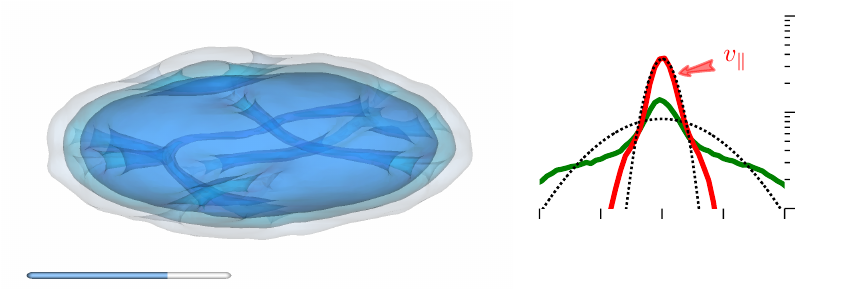}\\  
  \tmp{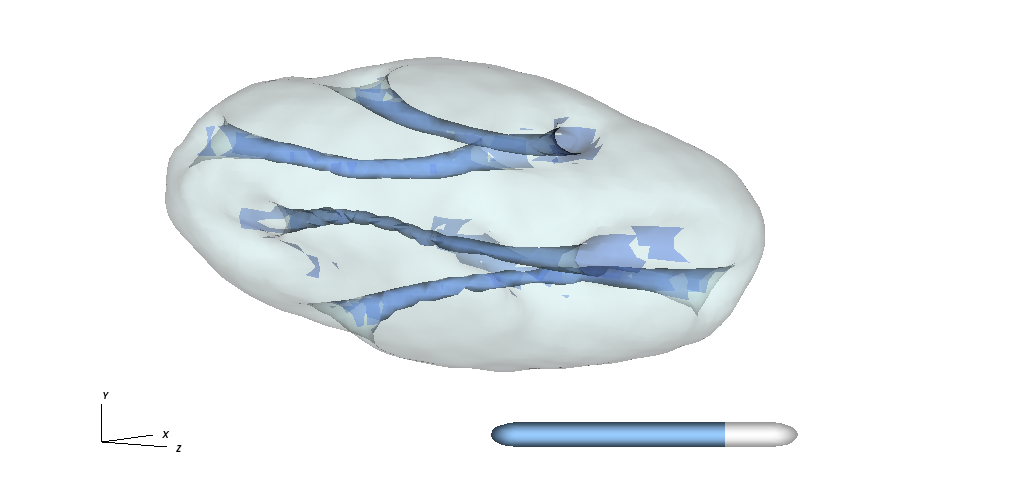}{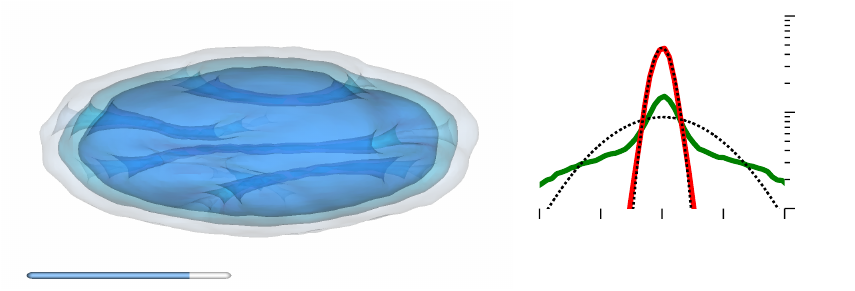}
  \let\tmp\relax
  \caption{\label{fig:qt_gen} Generation of quantum turbulence
    with a phase imprint of the vortex lattice~\cite{Wlazlowski:2015}.  In the
    left column consecutive frames show a vortex lattice with the knife edge
    dividing the cloud (top frame) and the decay of turbulent motion. In the
    right column we show the corresponding \gls{PDF} for longitudinal $v_{||}$
    and transversal $v_\perp$
    components of collective velocity (in arbitrary units). Dotted lines show
    the Gaussian best fit to the data.}
\end{figure*}

Attempts were made to simulate the MIT experiments using the same \gls{GPE}
tuned to the \gls{UFG} equation of state~\cite{Ancilotto:2012} that explained
the shock-wave experiment~\cite{Joseph:2011}.  While these kind of 
simulations were able to reproduce
some aspects of the vortex dynamics, all of our attempts to generate a stable
vortex ring out of a domain wall using only a \gls{GPE} treatment failed. The
main reason seems to be that the \gls{GPE} lacks any mechanism other than
phonon emission for dissipating excitations, apart from the fact that there is
no way to distinguish between the normal density and anomalous density 
in such an approach.  
This lack of dissipation is related to the inability of the pure
\gls{GPE} to crystalize vortex lattices.  We are not aware of any successful
result to stabilize vortex rings in the \gls{GPE} without \textit{ad hoc}
mechanisms,  introduced to remove energy from the
system~\cite{Scherpelz:2014}.  It is possible that dynamics in fermionic
superfluids may admit a simpler description in terms of \gls{GPE}-like
hydrodynamics, but direct comparison with the \gls{TDSLDA} demonstrates that
additional dissipation is required~\cite{Forbes:2012b}, and any such theory
needs a careful and thorough validation against the \gls{TDSLDA} and experiment,
which has yet to be performed.

Interestingly, both of these effects seem to be properly characterized by the
\gls{TDSLDA} treatment, even though it formally conserves energy.  In
particular, the \gls{TDSLDA} allows for pair-breaking effects, and having the
fermionic degrees of freedom allows the theory to dissipate excess energy along
the vortices such that vortex lattices crystalize~\cite{Bulgac:2011b} and the
vortex rings stabilizes~\cite{Bulgac:2013d,Wlazlowski:2015}.  In a \gls{GPE}
simulation of a recent experiment on the Josephson effect in fermionic
condensates across the \gls{BCS}-\gls{BEC} region the authors also failed to
reproduce the generation of vortices at high values of the current across the
junction in the neighborhood of the Feshbach resonance~\cite{Valtolina:2015},
which points to the need to simulate such an experiment within \gls{TDSLDA}.
 
The availability of a reliable theory to address fine details and make quite
accurate predictions is not available for liquid helium, and together with the
ability of experimenters to visualize in detail the spatial vortex
configurations and their time
evolution~\cite{Ku:2015, Serafini:2015} makes fermionic
cold-atom systems quite unique. One can now also mix cold Bose and Fermi
systems~\cite{Ferrier-Barbut:2014,Roy:2016,Yao:2016} and can hopefully study
in the future in detail both theoretically and experimentally their
entrainment, a phenomenon predicted a long time ago by Andreev and
Bashkin~\cite{Andreev:1975,Volovik:1975a}, and so far not yet put in evidence in
any other system. The Andreev-Bashkin entrainment can be relevant to the
neutron star
physics~\cite{Vardanyan:1981,Alpar:1984x,Gusakov:2005,Chamel:2006,Chamel:2008},
where the two superfluids, the superfluid neutrons and the superconducting
protons and their respective normal components, and the electrons in a normal phase, and their currents and vortices can
entrain each other and where experiments or direct observations cannot be
performed. Moreover, as in any other superfluid in rotation, where a massive
number of quantized vortices are formed, which also have to exists in a
background of a Coulomb lattice of nuclei immersed in a neutron superfluid, the
pinning and unpinning of vortices occur, the vortices deform, and during their
evolution they can cross and recombine, and quantum turbulence
emerges~\cite{Peralta:2006}.

The cold Fermi gases studied in the neighborhood of the Feshbach
resonance have several ``unfair'' advantages over other physical
systems.  One can easily manipulate the magnitude and even the sign of
the interaction and thus be able to confront theory in experiment in a
much wider range of parameters than otherwise one can envision. The
number of particles, their density, the temperature, the shape of the
confining potentials, a multitude of external probes and stirrers can
be rather easily experimentally realized and theoretically
simulated. Since the interaction is strong in the Fermi gases,
the coherence length is comparable to the average interparticle
separation, and the gases are dilute, the vortex-line separation can becomes
comparable to the inter-particle separation for modest rotation rates.
For example in \gls{TDSLDA} simulations presented in
Refs.~\cite{Bulgac:2011b, Wlazlowski:2015},
the average separation between vortices (forming the vortex) lattice was
about 3 times average inter-particle distance.
(see also top panel of Fig.~\ref{fig:qt_gen}).

%&&&&&&&&&&&&&&&&&&&&&&&&&&&&&&&&&&&&&&&&&&&&&&&&&&
\begin{figure*}[t]
   \def\tmp#1{\includegraphics[width=0.2\textwidth, trim=30 30 100 60, clip]{#1}}
   \tmp{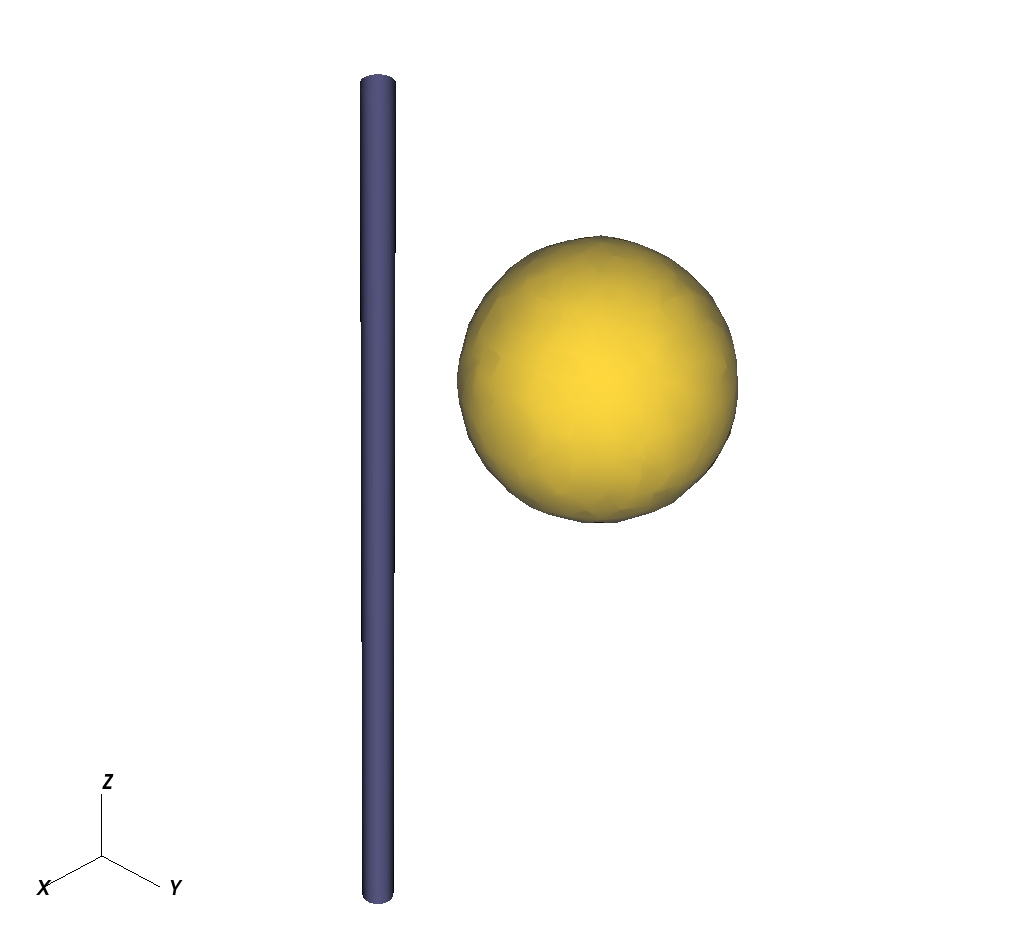}%
   \tmp{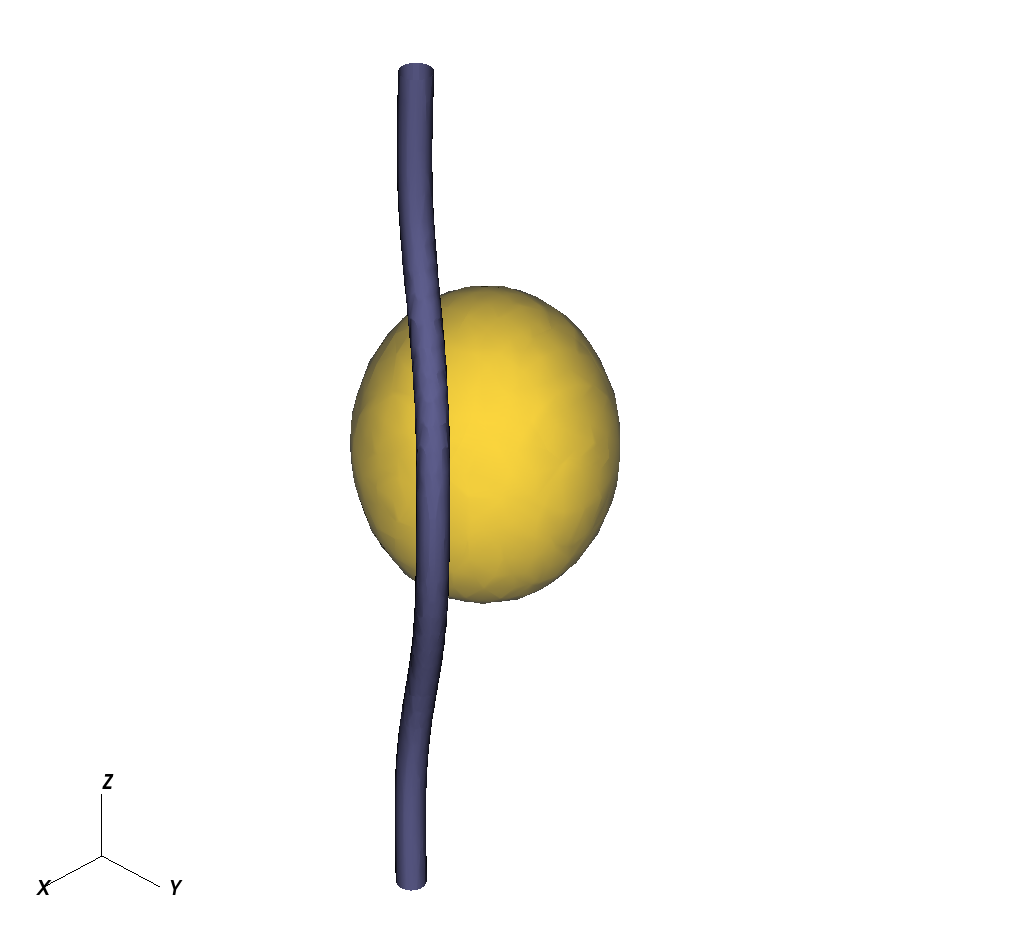}%
   \tmp{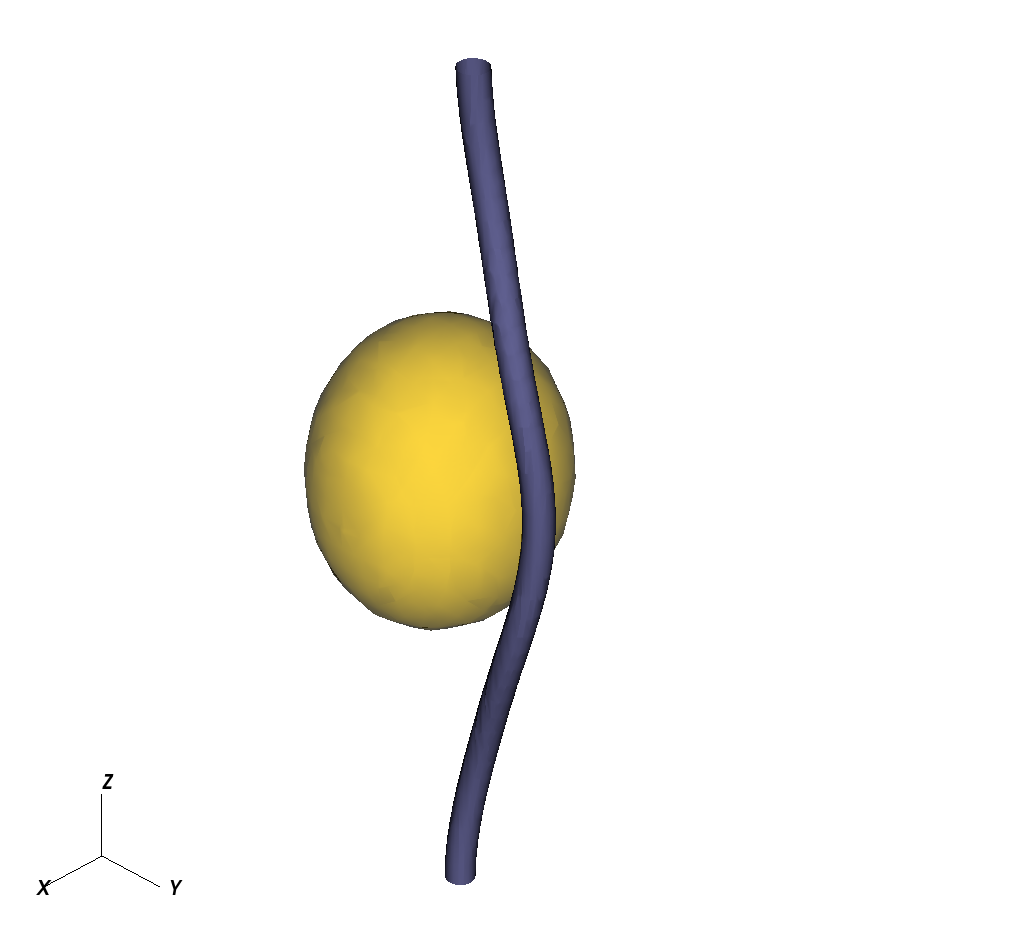}%
   \tmp{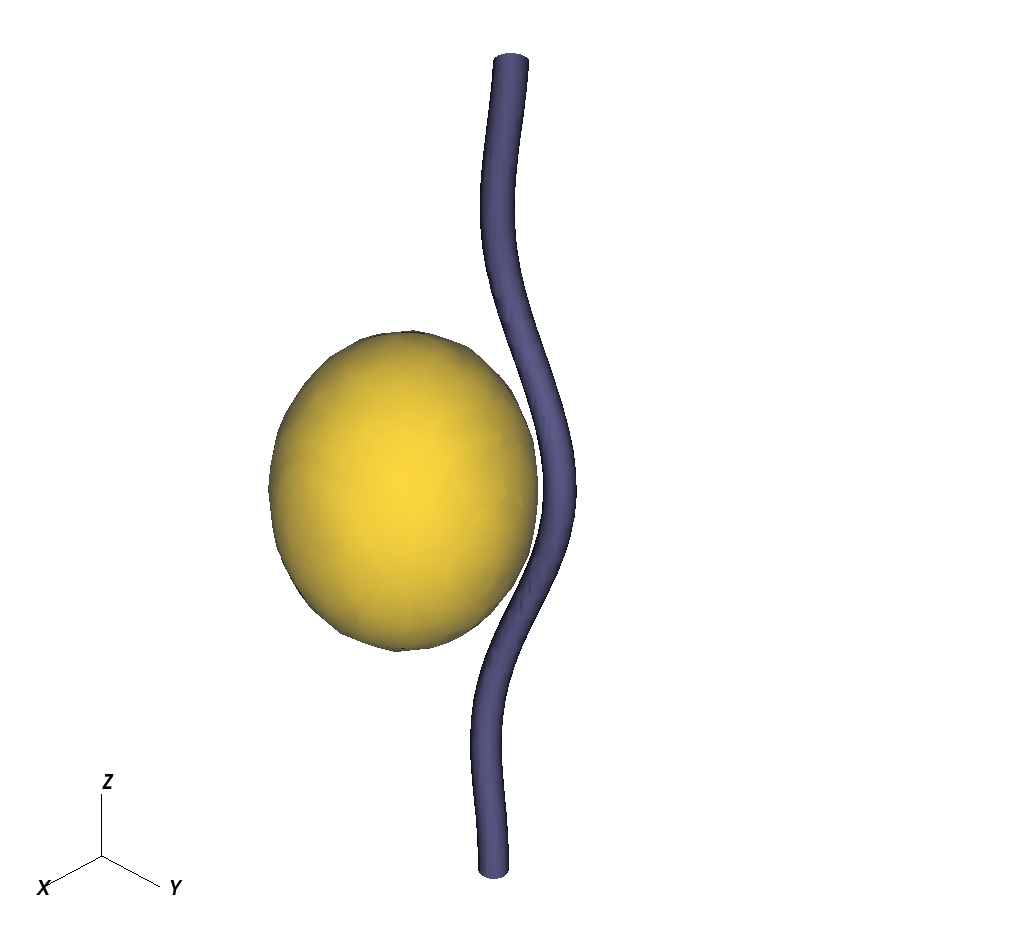}\\
   \def\tmp#1{\includegraphics[width=0.2\textwidth, trim=30 25 250 130, clip]{#1}}
   \tmp{vi0006.png}%
   \tmp{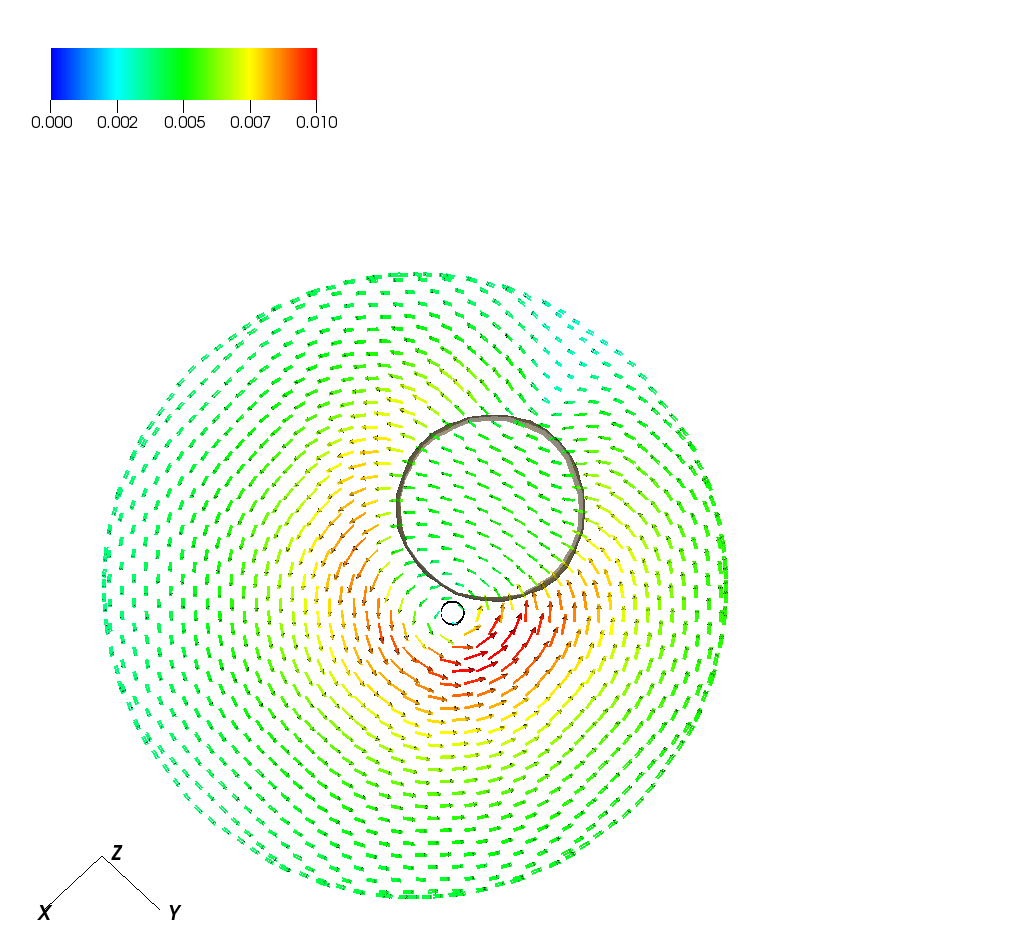}%
   \tmp{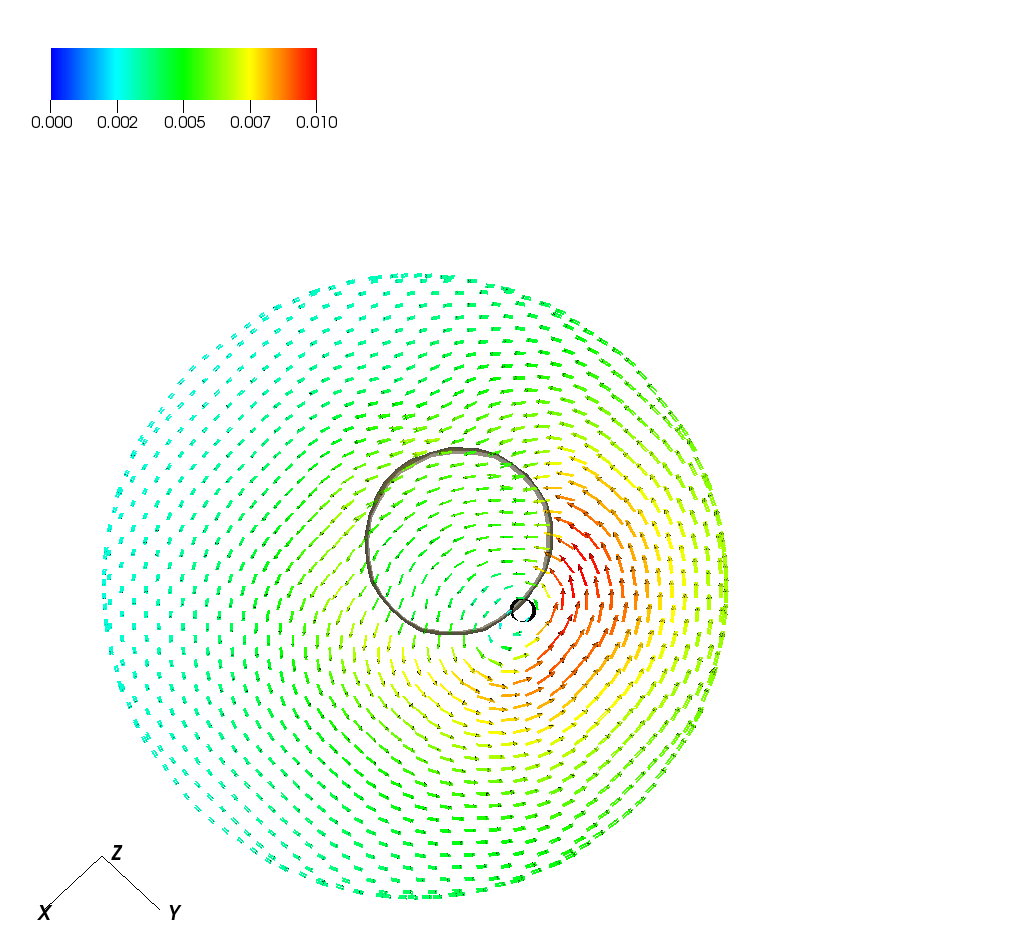}%
   \tmp{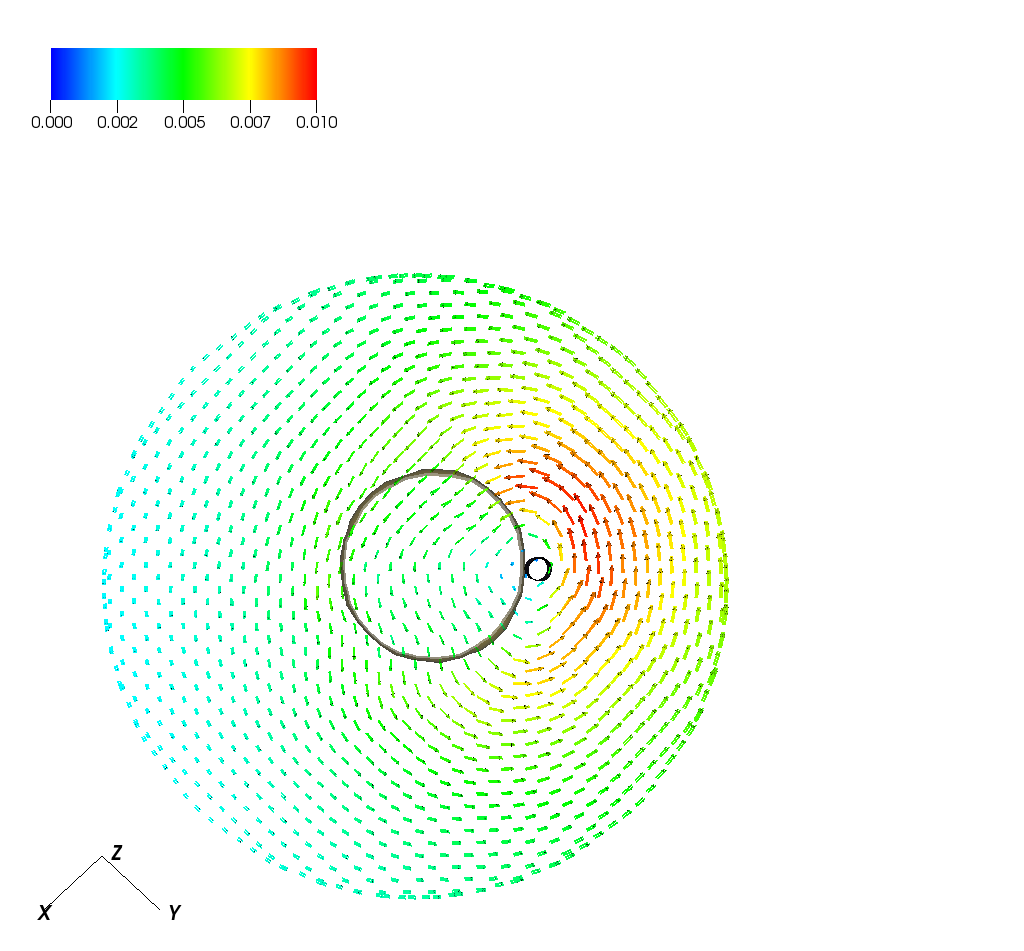}\\
   \let\tmp\relax
   \vspace{2mm}%
   \includegraphics[width=0.35\textwidth, trim=30 800 600 80, clip]{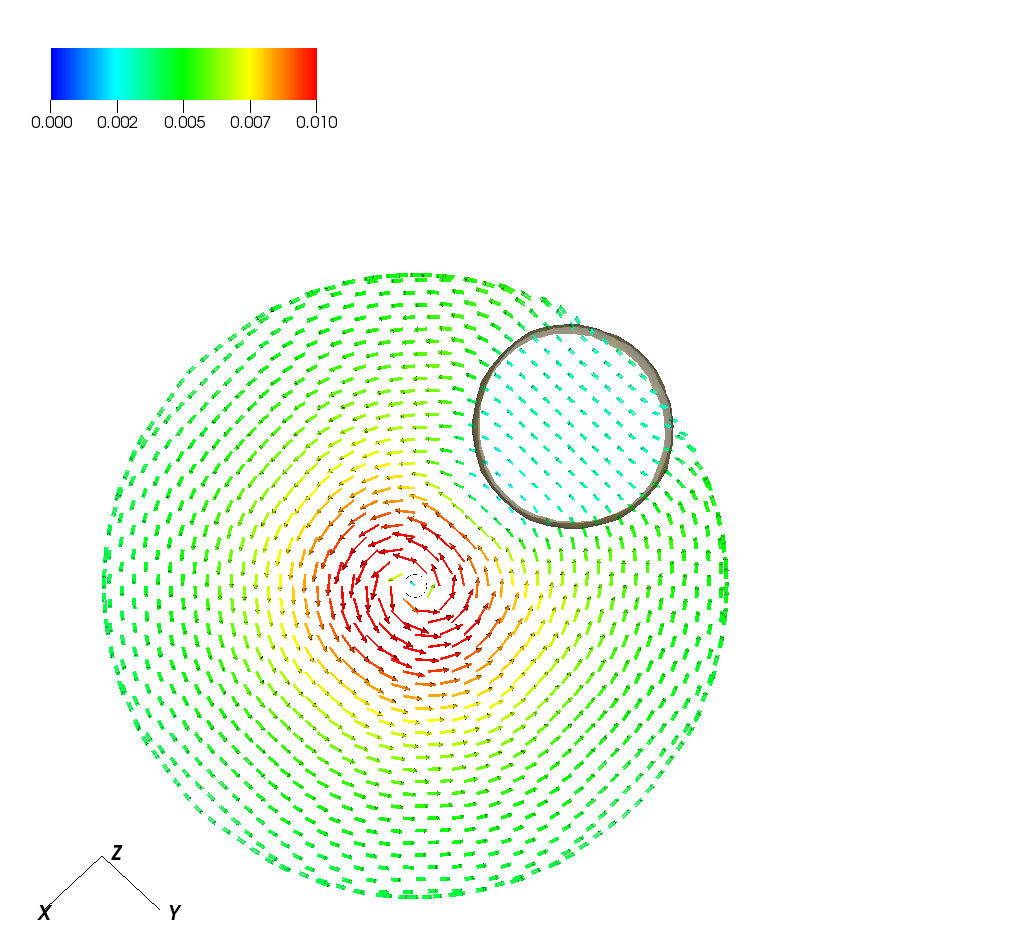}%
   \hspace{0.45\textwidth}
   \vspace{-1.4mm}
   \caption{Snapshots of the system ($n=0.031\,\text{fm}^{-3}$)
     for times $(0, 10, 14, 17)\times 10^3$~fm/$c$.
     Top line: A blue tube represents position of the vortex core. A yellow
     contour indicates the surface of the nuclear impurity (where proton
     density drops to value $0.005\,\text{fm}^{-3}$).
     The nucleus is moving along $x$-axis
     at a constant speed much smaller than the critical velocity
     $v= 0.001c \ll v_c$,
     due to an applied external force that is constant in space, and at each
     instance of time, exactly balances the force exerted by the vortex.  Notice the
     strong deformations underwent by both the vortex line and the nucleus
     during the evolution.  Bottom line: Velocity field of neutrons is
     presented by vectors (color indicates its magnitude) for a plane that
     crosses nucleus center.  Location of the nucleus and vortex core is also
     displayed.  The velocity field is affected by presence of nucleus. It is
     about 2 times smaller inside nucleus (arrows are green) than on the
     opposite side of the vortex core (arrows are red).  }\vspace{-4mm}
   \label{FIG:shape}
\end{figure*}

In Fig.~\ref{fig:qt_gen} we show an example of how one can generate a large
number of quantized vortices in a rather small cloud with only 1410 fermions,
together with a domain wall, which subsequently evolve, cross and recombine,
leading to manifestly non-equilibrium \glspl{PDF} of the particle
velocities~\cite{Wlazlowski:2015}.  We start with the ground state of a cloud
cut in half with a knife-edge potential and subsequently we then stir the
system with two circulating laser beams parallel to the long axis of the trap.
Once a vortex lattice is generated, we imprint a $\pi$
phase shift between the the two halves of the cloud.  Just before removing the
knife-edge, we introduce a slight tilt to speed the formation of a vortex
tangle.  After the knife-edge is removed, the vortex lines twist, cross, and
reconnect, and vortices untie in a manner qualitatively in agreement with a
recent theoretical analysis of vortex knots~\cite{Kleckner:2016}.
%  (Note: knots have now been realized experimentally~\cite{Hall:2016}).
From the velocity \gls{PDF} one sees a clear departure from the equilibrium
Gaussian behavior as the tangle evolves -- a hallmark of quantum
turbulence~\cite{White:2010, Barenghi:2001}.  Eventually the system relaxes to
a vortex lattice and equilibrates in $v_\parallel$.
Somewhat similar velocity \glspl{PDF} are seen in the theoretical studies of
dilute Bose gases~\cite{White:2010} and in the phenomenological filament model
of the crossing-recombination vortex line dynamics~\cite{Adachi:2011}.

A question of crucial importance in nuclear astrophysics is the pinning
mechanism of quantized vortices in the neutron star crust.  It was conjectured
almost four decades ago by Anderson and Itoh~\cite{Anderson:1975} that a
neutron star glitch, a sudden spin-up of the rotational frequency, is caused by
a catastrophic unpinning of a huge number of quantized vortices in the neutron
superfluid from the Coulomb lattice formed by the nuclei immersed in this
fluid.  It has been a matter of debate ever since whether a quantized vortex is
pinned to nuclei forming the lattice or interstitially, therefore whether a
vortex is attracted or repelled by a nucleus.  The interaction of a quantized
vortex with a nucleus immersed in a neutron superfluid is an extremely
difficult problem to solve theoretically and a direct observation is
impossible, and over four decades theorists produced contradictory predictions.
At the same time it is known that the properties of dilute neutron matter are
very similar to the properties of a \gls{UFG}. By applying the \gls{TDSLDA} and
using an accurate nuclear \gls{EDF} it was shown
recently~\cite{Wlazlowski:2016} that a vortex and a nucleus repel always, by
studying the dynamics of such a system and extracting the force between a
vortex line and a nucleus.  The interaction is rather complex, the nucleus and
the vortex line deform during their evolution. The situation illustrated in
Fig.~\ref{FIG:shape}, apart from the presence of a vortex, is identical to the
experimental setups realized in mixtures of Fermi and Bose systems
already~\cite{Ferrier-Barbut:2014, Roy:2016}, where spatially the Bose system
is of smaller size than the Fermi system.  This kind of experiments can be
easily performed in the presence of one or even many vortices in the presence
of one or even many ``impurities,'' and the motion of such impurities can be
also controlled. Performing and studying such systems will definitely shed a
lot of light on the vortex-pinning mechanism and the dynamics in neutron star
crust.

\glsunset{LOFF} \glsunset{FFLO} Another degree of freedom to be exploited in
future studies of quantum turbulence is the spin polarization of the Fermi
systems. Clogston~\cite{Clogston:1962} and
Chandrasekhar~\cite{Chandrasekhar:1962} noted the normal phase competes with
the \gls{BCS} superfluidity phase in the case of spin polarized Fermi systems.
It was predicted a long time ago that the $s$-wave
pairing mechanism is modified by the emergence of the so called
\gls{LO}~\cite{LO:1965} and \gls{FF}~\cite{FF:1964} phase (\gls{LOFF} or
\gls{FFLO}), when the two fermions in the Cooper pair couple to a non-zero
total momentum and the pairing gap oscillates in space. The equation of state
of the polarized \gls{UFG} has been extracted independently
experimentally~\cite{shin-2008} and within the
\gls{SLDA}~\cite{BF:2008,Forbes:2008}, where it was also shown that the
\gls{LOFF} phase exists at unitarity for a wide range of spin polarization. The
studies of the structure and dynamics of vortices, and particularly the
possibility and features of the quantum turbulence, in this regime in the
\gls{UFG} still awaits its implementation. It was shown recently the the
Clogston-Chandrasekhar critical polarization for fermionic superfluidity is
enhanced in Fermi-Bose systems~\cite{Ozawa:2014}, making such systems even more
promising to study in non-equilibrium.

Cold Fermi gases around the Feshbach resonance have another peculiar feature.
Systems with fewer than about a million atoms can have quantum turbulence below the critical temperature.
However, this number is too small to sustain large-scale flow and support classical turbulence~\cite{Wlazlowski:2015a}.
In this respect these systems are different from traditional liquid \ce{^4He} experiments at finite temperatures, where both normal and superfluid systems could become turbulent.
Classical turbulence in 3D systems is achieved for values of the Reynolds number of the order of $10^4$ or higher~\cite{Lamb:1945,LL6:1966}.
In the \gls{UFG} quantized vortices, and therefore quantum turbulence, can exist in clouds with as little as 500-1000 fermions and flow velocities $v\approx 0.7v_F$, see supplemental material in Ref.~\cite{Bulgac:2011b}.
Interestingly, due to the compressibility of these gasses, this velocity can be larger than the average Landau critical velocity $v_c\approx 0.4v_F$.
The largest cold atomic clouds created so far in the laboratory have about $10^6$ atoms and a Reynolds number $\Re \approx 600$.
The Reynolds number scales with particle number roughly as $\Re = nmLv/\eta \sim N^{1/3}$ at constant atom number density $n$, where $m$ is the mass of the atom, $L$ the characteristic size of the system, and $\eta$ the shear viscosity.
Increasing the flow velocity to values much higher than the Fermi velocity or critical velocity will render the system normal, with properties close those of a free Fermi gas.

\section*{Conclusion}
Superfluid Fermi gases provide a new forum in which to study quantum
turbulence. As with cold atomic \glspl{BEC}, one has exquisite experimental
control with the ability to excite and observe topological defects such as
domain walls and vortices that lie at the heart of quantum turbulence.  Unlike
\glspl{BEC}, the unitary Fermi gas is strongly interacting and demonstrates an
extremely high vortex line density.  Consequently, turbulence can be realized
in a new regime: namely in small systems where classical turbulence is
suppressed.  To date, small numbers of topological defects have been formed and
imaged in cold Fermi gases and there appears to be no impediment to realizing
turbulence.  Current imaging processes, however, are generally destructive and
require expansion which can impede their interpretation.  Some questions
concerting phenomena related to quantum turbulence may therefore require
improved non-destructive imaging techniques, perhaps using multi-component
systems as has been done with \glspl{BEC}~\cite{Anderson:2000} or using some
sort of tracer as done with \ce{He}~\cite{Paoletti:2011, Fonda:2012, Fonda:2014, La-Mantia:2014}. Particularly useful
would be the development of imaging techniques that allow for the extraction of
the velocity distribution in a turbulent system.

Although Fermi gases can realize turbulence in small systems with less than
1000 fermions, these are still beyond the reach of direct ab initio quantum
techniques like \gls{QMC}.  Validated dynamical models will therefore be
required.  In weakly interacting systems, \gls{GPE}-based theories and
\gls{ZNG}-like extensions work well, and similar techniques may apply for Fermi
gases, but the most accurate approach that remains tractable for strongly
interacting systems appears to be superfluid \glspl{DFT} such as the \gls{SLDA}
which has been validated at the percent level for static systems and
qualitatively validated against dynamical experiments.  Precision measurements
of quantitative dynamics in relatively small systems ($\lesssim 500$
particles) is now required to quantitatively benchmark the dynamical component
of these theories.  The extension of \gls{TDSLDA} to include fluctuations will
also be needed. Above the critical temperature, in the pseudogap regime, where
both dimers and atoms coexist, a new kinetic or hydrodynamic approach would be
desirable as well.  Benchmarking these theories is critical for other branches
of physics.  In particular, dilute neutron matter in neutron stars admits no
experimental access but is described by the Fermi gas near the Feshbach
resonance with reasonable accuracy.  Validated dynamical theories for Fermi
superfluids are thus critical to resolve outstanding astrophysical problems
such as the origin of pulsar glitches, which are thought to originate from
vortex dynamics in the neutron superfluid, providing an astrophysical venue in
quantum turbulence may play a key role.

Finally, strongly interactive Fermi superfluids and superfluid mixtures exhibit
exotic behaviors such as supersolid \gls{LOFF} phases, and Andreev-Bashkin
entrainment, which have only begun to be explored.  The next decade has the 
great potential to be an exciting time for studying quantum turbulence.

\begin{acknowledgments}
 We are grateful to our colleagues and collaborators over the years
G.F. Bertsch, Y.-L. Luo, P. Magierski, K.J.~Roche, A.~Sedrakian,
R. Sharma, S. Yoon, and Y. Yu for many helpful discussions and comments.
This work was supported by
the Polish National Science Center (NCN) under Contracts
No. UMO-2013/08/A/ST3/00708 and No. UMO-2014/13/D/ST3/01940.
Calculations have been performed: at the OLCF Titan - resources of the
Oak Ridge Leadership Computing Facility, which is a DOE Office of
Science User Facility supported under Contract DE-AC05-00OR22725; at
NERSC Edison - resources of the National Energy Research Scientific
computing Center, which is supported by the Office of Science of the
U.S. Department of Energy under Contract No. DE-AC02-05CH11231; at
HA-PACS (PACS-VIII) system - resources provided by Interdisciplinary
Computational Science Program in Center for Computational Sciences,
University of Tsukuba. This work was supported in part by U.S. DOE
Office of Science Grant DE-FG02-97ER41014, and a WSU New Faculty Seed
Grant.
\end{acknowledgments}

%\bibliography{local,master}

\end{document}